\documentstyle[12pt]{article}
\begin{document}
\thispagestyle{empty}
\setlength{\baselineskip} {4.0ex}
\newcommand{\p}{$\overline{p}$}
\newcommand{\n}{$\overline{n}$}
\newcommand{\N}{$\overline{N}$}
\newcommand{\D}{$\overline{d}$}
\def\ie{{\it i.e.} }
\def\etal{{\it et al.}}
\def\p{$\pi$ }
\def\r{$\rho$ }
\def\o{$\omega$ }
\def\e{$\eta$ }
\def\s{$\sigma$ }
\def\ns{$N^*$ }
\def\en{$\eta$N }
\def\nn{$NN$ }
\def\nne{$NN \rightarrow NN \eta$ }
\def\ppe{$pp \rightarrow pp \eta$ }
\def\npe{$np \rightarrow np \eta$ }
\def\ppp{$pp \rightarrow pp \pi^0$ }
\def\de{$np \rightarrow d \eta$ }
\def\sa{$S/A$ }
\def\pt{$p_T$ }
\def\et{$E_T$ }
\def\fm3{fm$^3$}
\def\fmm3{fm$^{-3}$}
\def\fsi{$FSI\ $}
\def\isi{$ISI\ $}
\def\s11{$S11\ $}
\def\pr{Phys.\ Rev.\ }
\def\pra{Phys.\ Rev.\ }
\def\prb{Phys.\ Rev.\ }
\def\prc{Phys.\ Rev.\ }
\def\prd{Phys.\ Rev.\ }
\def\prl{Phys.\ Rev.\ Lett.\ }
\def\pl{Phys.\ Lett.\ }
\def\np{Nucl.\ Phys.\ }
\def\ap{Ann.\ Phys.\ (N.Y.) }
\def\prep{Phys.\ Rep.\ }
\def\jp{J.\ Phys.\ }
\def\zp{Z.\ Phys.\ }
\def\mpl{Mod.\ Phys.\ Lett.\ }
\def\rmp{Rev.\ Mod.\ Phys.\ }
\def\sjnp{Sov.\ J.\ Nucl.\ Phys.\ }
\title{On The $pp \rightarrow pp \eta (\eta')$ Reactions Near Threshold}  
\author{ E.Gedalin\thanks{gedal@bgumail.bgu.ac.il},
A.Moalem\thanks{moalem @bgumail.bgu.ac.il} and L.Razdolskaja\thanks
{ljuba@bgumail.bgu.ac.il}}                                    
\date{Department of Physics \\
 Ben-Gurion University of the Negev\\
Beer-Sheva, 84105,  Israel\\}

\maketitle
\begin{abstract}

The production rate for $\eta'$ in $pp \rightarrow pp \eta'$ at rest
is calculated in a covariant one boson exchange model, 
previously applied to study $\pi^0$ and $\eta$ production in NN 
collisions. The transition amplitudes for the elementary 
$BN \rightarrow \eta' N$ processes with $B$ being the meson exchanged 
( $B = \pi , \sigma , \eta , \rho , \omega$ and $\delta$) are taken 
to be the sum of s and  u channels with a nucleon in the intermediate 
states, 
and a $\delta$ meson pole in a t-channel. The couplings of the $\eta'$
to hadrons are a factor 0.437 weaker than the respective 
$\eta$-hadron couplings, as suggested by a quark model and a singlet-octet
mixing angle $\theta = -23^o$. The model reproduces near threshold
cross sections for the quasielastic processes
$\pi^- p \rightarrow n \eta (\eta') $  
and $pp \rightarrow pp \eta (\eta')$ reactions. 

\end{abstract}
\vfill
\pagebreak
\section{Introduction}
\bigskip
The production of $\eta'$ mesons in proton-proton collisions near 
threshold has been reported recently by the SPESIII group
at Saturne\cite{hibou98} and  by COSY-J$\ddot u$lich\cite{moskal98}. 
The $\eta'$ is observed as a missing mass peak by detecting  
both final protons in a magnetic spectrometer. The total cross section 
of the $pp \rightarrow pp \eta'$ reaction  is found to be a factor 
$\approx (m_{\pi}/m_{\eta '})^2$ smaller than for the corresponding 
$pp \rightarrow pp \pi^0$, 
indicating  a similar production mechanisms for $\eta '$ and $\pi^0$.
The observation that the $\eta$ production rate is nearly as large as 
$\pi^0$ production rate is attributed to a dominant contribution
from resonant production via virtual excitation of the 
N$^*$ (1535 MeV) S11 nucleon isobar.

Various phenomenological one-boson-exchange (OBE) models for these 
processes have achieved impressive descriptions of extensive data
from Saclay\cite{plouin90,bergdolt93,chiavassa94}, 
Indiana\cite{meyer92}, 
and Uppsala 
\cite{schuberth95,bondar95,calen96,haggstrom97,calen97}. 
Particularly, a relativistic covariant OBE model, reproduces consistently
the cross section data (both scale and energy dependence) for the
$pp \rightarrow pp \pi^0$ and $pp \rightarrow pp \eta$ 
reactions\cite{gedalin98,gedalin981}. The 
main objective of the present note is to consider $\eta'$ production
by applying a slightly generalized model.
As in Ref.\cite{gedalin98,gedalin981} we assume a reaction mechanism as
depicted in Fig. 1, where a virtual boson B (B= $\pi , \eta , \sigma , \rho ,
\omega , \delta )$ created on one of the incoming protons is converted
into a pseudoscalar meson P ($\pi ,\eta , \eta'$) on the second via a 
$BN \rightarrow PN$ 
conversion process. The half off mass-shell amplitudes for these
conversion processes, hereafter denoted by $T_{BN \rightarrow PN}$,
are taken to be the sum of three pole terms corresponding to s, u, and 
t-channels (see Fig. 2). 
The latter accounts for meson production 
occurring on internal lines such as $\sigma$ and $\delta$ meson lines,
and requires knowledge about three meson legs vertices like  
$\sigma \pi \pi$, $\delta \eta \pi$ and $\delta \eta '\pi$ 
vertices. Such mechanisms are
considered in Ref. \cite{gedalin981}, where an effective 
$\sigma$-meson in a t channel, strongly enhanced due to
offshellness, found to play a prominent role for $\pi^0$ 
production in the $pp \rightarrow pp \pi^0$ reaction at threshold.
For $\eta '$ production most relevant is  a $\delta \pi $ exchange.
Generally, both nucleon and nucleon isobar 
excitations intermediate states may contribute to  diagrams 2a and
2b. Such is the case for $\eta$ production, where production via 
exciation, propagation and subsequent decay of the N$^*$ (1535 MeV)
S11 isobar into a $\eta$N pair play a significant role in the process. 
We shall demonstrate below that 
cross section data for $\eta '$ production can be
explained without a resonant production term. 

Much of the model success to explain meson production data 
in NN collisions depends on how well the half off mass shell amplitudes 
for the elementary conversion  processes $BN \rightarrow PN$ 
are calculated. Unfortunately, there are no direct measurements which
link directly to the off mass shell behavior of these amplitudes. 
In comparison with the $pp \rightarrow pp \eta$ reaction, 
the momentum transfer 
in $pp \rightarrow pp \eta'$ is nearly twice as large and it 
would be of interest to verify that the off shell behavior 
presumed reproduces $\eta'$ production rate as well.
There are some near threshold cross section data 
for the $\pi N \rightarrow N \eta (\eta')$ reactions\cite{binnie73}. 
Here as for production in NN collisions the cross sections depart 
by more than an order of magnitude\cite{binnie73}.
It is to be shown that taken on mass shell, the amplitudes 
$T_{\pi N \rightarrow  N\eta (\eta')}$   
reproduce the cross sections for the quasielastic processes 
$\pi^0 p \rightarrow p \eta$ and 
$\pi^0 p \rightarrow p \eta'$ at their respective thresholds.

\section{Theoretical perspective} 
\bigskip
To account for the production of $\eta'$ meson, we slightly
generalize the Lagrangian of Refs. \cite{gedalin98,gedalin981}, 
by including a Lagrangian density,
\begin{equation}
L_{\eta'NN} = \frac {f_{\eta' NN}}{m_{\eta'}}\bar{N}\gamma^5 
       \gamma^{\mu}\partial_{\mu}\eta'  N ~,
\label{lagran} 
\end{equation}
where N and $\eta'$ represent  the fields of a nucleon and
an $\eta'$ meson.

Further more, in order to allow $\eta'$ formation on
an internal $\delta$-meson line we write the $\delta \eta' \pi$ vertex in 
a rather general form\cite{gedalin981},
\begin{equation}
V_{\delta \eta' \pi} (k^2,q^2,(k-q)^2)= m_{\delta} \left( g_{0\eta'} 
    + g_{1\eta'} \frac {k^2}{m_{\delta}^2}
    + g_{2\eta'} \frac {q^2}{m_{\delta}^2}
    + g_{3\eta'} \frac {(k-q)^2}{m_{\delta}^2}\right)~.
\label{vertex}
\end{equation}
This expression is similar to the $\delta \eta \pi$  vertex 
used in Ref. \cite{gedalin981} for $\pi^0$ 
(or $\eta$) production, where the vertex constants $g_{i\eta}; i=0 - 3$ 
were deduced from the partial decay width for 
$\delta \rightarrow \pi \eta$,
and by applying the three Adler's consistency conditions to the
$T_{\pi^0 p \rightarrow \eta p}$ amplitude\cite{gedalin981}.

There is no direct experimental information about the strength of the
$\eta'$ couplings. However, being the heaviest member of 
the ground state pseudoscalar meson nonet, the $\eta'$ couplings  
can be related to those of the $\eta$. In the quark model the $\eta$NN and 
$\eta'$NN couplings are\cite{pdg}, 
\begin{eqnarray}
 & & g_{\eta NN} = g_8 \cos {\theta} - g_1 \sin {\theta}~, \nonumber \\
 & & g_{\eta' NN} = g_8 \sin {\theta} + g_1 \cos {\theta}~,
\end{eqnarray}
with $\theta, g_1 , g_8$  being the mixing angle, singlet and 
octet couplings, respectively. By making the simplifying assumption 
$g_8 = g_1$, and taking a linear mixing angle\cite{pdg}, $\theta = -23^o$, 
one obtains,
\begin{equation}
\frac {g_{\eta' NN}}{g_{\eta NN}} = 0.437~.
\label{ratio}
\end{equation}
Similarly, the $\delta \eta \pi$ and $\delta \eta' \pi$ 
vertex constants scale also
according to Eq. \ref{ratio}.  With the relevant $\eta$ couplings taken 
from Ref.\cite{gedalin981} and disregarding uncertainties in $\theta$
one obtains
\begin{eqnarray}
& & g_{\eta' NN} = 2.68,~~~ g_{0\eta'} = 0.02 \pm 0.01,~~~ 
g_{1\eta'} = -2.6 \pm 0.20, \nonumber \\
& & g_{2\eta'} = -1.51 \pm 0.12,~~~ g_{3\eta'} = 1.44 \pm 0.11~.
\end{eqnarray}

The  $\eta$ meson couples rather strongly to the N$^*$ (1535 MeV)
S11 and to a lesser extent to the N$^*$ (1710 MeV) P11 resonances.
Particularly, the resonance mass almost coincides with
the mass of a $\eta$ N pair so that contributions from graphs 
2a and 2b with nucleon isobar excitations become prominent near the 
$\eta$ production threshold. 
There are no evidence for strong couplings of the $\eta'$ to baryon 
resonances.  The decays of the N$^*$ (1535 MeV) S11 
and N$^*$ (1710 MeV) P11 resonances into a free $\eta '$N pair are
not accessible either. It is then  reasonable to assume that 
resonant $\eta '$   production terms are small and can be neglected.
With this in mind and following
Refs.\cite{gedalin98,gedalin981} the amplitudes for the
$\pi^0 p \rightarrow \eta p$ and $\pi^0 p \rightarrow \eta' p$ are, 
\begin{eqnarray}
 & & T_{ \pi^0 p \rightarrow \eta p } = \nonumber\\
 & & -i g_{\eta NN^*} g_{\pi NN^*} f_{\eta} (k) f_{\pi} (q) 
\bar {u} (p') \left[ \frac {1}{M_R - \sqrt {s} + i \Gamma /2} + 
\frac {1}{M_R - \sqrt {u} + i \Gamma /2}\right] u(p)\nonumber \\
 & & i\frac {2M f_{\pi NN}}{m_{\pi}} \frac {2M f_{\eta NN}}{m_{\eta}}
     f_{\eta}(k) f_{\pi}(q)
{\bar u} (p') \left[k\! \! \! / \left(\frac {1}{s-M^2} - 
\frac {1}{u-M^2}\right) - \frac {1}{M}\right] u(p) + \nonumber \\
 & & i g_{\delta NN} \frac {f_{\delta}(q - k)} {(q - k)^2 - m_{\delta}^2}
V_{\delta \eta \pi}(k,q)
{\bar u}(p') u(p)~,
\label{vr:dep1}
\end{eqnarray}
\begin{eqnarray}
 & & T_{\pi^0 p \rightarrow \eta' p} = \nonumber\\
 & & i\frac {2M f_{\pi NN}}{m_{\pi}} \frac {2M f_{\eta' NN}}{m_{\eta'}}
     f_{\eta'}(q) f_{\pi}(k)
{\bar u} (p') \left[k\! \! \! / \left(\frac {1}{s-M^2} - 
\frac {1}{u-M^2}\right) - \frac {1}{M}\right] u(p) + \nonumber \\
 & & i g_{\delta NN} \frac {f_{\delta}(q - k)} {(q - k)^2 - m_{\delta}^2}
V_{\delta \eta' \pi}(k,q)
{\bar u}(p') u(p)~,
\label{vr:dep2}
\end{eqnarray}
where $M_R$, $\Gamma$ represent the mass and width of the 
N$^*$ (1535 MeV) S11 resonance; $M$ and $m_B$ the mass of a nucleon 
and a boson $B$; $f_B$ a source form factor parametrized in the usual 
form\cite{machleidt89},
\begin{equation}
f_b = \frac {\Lambda_B^2 - m_b^2}{\Lambda_B^2 - q^2}~.
\end{equation}
It is easy to trace in the expressions above 
the contribution from s, u and t channels. The first two terms in 
Eq. \ref{vr:dep1} describe resonant production via nucleon isobar
excitations, while the last term in both of these expressions 
stands for contributions from a $\delta$ meson t pole. 
The other meson conversion amplitudes and numerical details
of the calculations are given in Refs.\cite{gedalin98,gedalin981}
and shall not be repeated here.

We may now use Eqns. \ref{vr:dep1} and \ref{vr:dep2} 
to evaluate the cross 
sections for the quasielastic processes $\pi^0 p \rightarrow \eta p$   
and $\pi^0 p \rightarrow \eta' p$.  
At threshold these expressions predict for the amplitude squared :
$|f(\pi^0 p \rightarrow \eta' p)|^2 = (15 \pm 7) \mu b/sr$ and 
$|f(\pi^0 p \rightarrow \eta p)|^2 = (346 \pm 40 )\mu b/sr$  
which are consistent with the experimental values\cite{binnie73} 
$|f(\pi^0 p \rightarrow \eta' p)|^2 = 10\pm 1 \mu b/sr$ and 
$|f(\pi^0 p \rightarrow \eta p)|^2 = 365 \pm 30 \mu b/sr$, respectively.
Predictions for $|f|^2$ are drawn in Figs. 3-4 versus the energy
available in the center of mass (CM) system. 
Due to mutual cancellation, the contribution from both of the s and u 
nucleon pole terms (drawn separately as dot-dashed curve)
is negligibly small for both processes. The
t pole terms are also of the same size though playing 
a different role in the two cases.
For the $\pi^0 p \rightarrow \eta p$ reaction the
resonant production  exceeds by far any
of the other contributions. 
The t pole term is relatively weak, becoming noticeable only  
through interference with the strong resonant production term.
In case of the $\pi^0 p \rightarrow \eta' p$ however, there is 
no strong resonance term and the t pole contribution determines 
the cross section almost solely.

We now turn to consider the cross sections for the
$pp \rightarrow pp \eta$ and $p p \rightarrow pp \eta'$ reactions.
Let us  call
\begin{equation}
\Pi_j = \frac {{\bf p}_j} {{E_j + M}}~,\\
\end{equation}
where  ${\bf p}_j$ and $E_j$ are three-momentum and total energy 
of the j-th nucleon. For the incoming particles in the CM system,
$\Pi_1 = - \Pi_2 = \Pi$. The total energy square 
and the energy available in the CM system are respectively,
$s = (p_1 + p_2)^2$ and $Q = \sqrt{s}- 2M - m_{\eta'} $.
There is only one isovector amplitude which determines the 
reaction cross section at rest, corresponding 
to a ${}^{33}P_0 \rightarrow {}^{31}S_0$ transition in the two 
nucleon system. We write this amplitude in the form
\begin{equation}
M_{11}(pp \rightarrow pp \eta') = M_{\pi} + M_{\eta} 
+ M_{\sigma} + M_{\delta} + M_{\rho} + M_{\omega}~,
\end{equation}
where $M_B$ represents the contribution from the exchange of a 
boson B. Following Ref.\cite{gedalin98},
\begin{eqnarray}
 & & M_{\pi}= i G_{\pi NN}  
g_{\pi NN}g_{\eta' NN}\Sigma_P \nonumber \\
 & & ~~-i f_{\pi NN} 
\left( \frac {2M}{m_{\pi}}\right)
\left( \frac {1}{m_{\pi}^2 - q^2}\right)f_{\pi }(q) g_{\delta NN}
\left( \frac {f_{\delta}(q)}{m_{\delta}-q^2}\right)
V_{\delta \eta' \pi} (k^2 = m_{\eta'}^2;q^2)~, \label{minpi} \\
 & & M_{\sigma} = i G_{\sigma NN} g_{\sigma NN}g_{\eta' NN}\Sigma_S~, 
\label{minsigma} \\
 & & M_{\eta} = i G_{\eta NN} 
g_{\eta NN}g_{\eta' NN}\Sigma_P~, \label{mineta} \\
 & & M_\rho = G_{\rho NN} 
g_{\rho NN}g_{\eta' NN}(\Sigma_{\rho}^{(1)} + 2 \Sigma_{\rho}^{(2)})~,~~
\label{minrhou} \\
 & & M_\omega = G_{\omega NN} 
g_{\omega NN}g_{\eta' NN}(\Sigma_{\omega}^{(1)} + 2 \Sigma_{\omega}^{(2)})~,
~~~~~\label{minomega} \\
 & &  M_{\delta} = i G_{\delta NN}  
g_{\delta NN}g_{\eta' NN}\Sigma_S~, \label{mindelta} 
\end{eqnarray}
where
\begin{eqnarray}
 & & G_{BNN}= g_{B NN}\frac{E + M}{M}\frac{1}{M(m_{\eta'}+Q)
+ m_B^2}f_B^2 (-M[m_{\eta'}+Q]) \Pi~,\\	 
 & & \Sigma_S =\frac{1}{M}\left(1 - \frac{5m_{\eta'}}
{2M+m_{\eta'}}\right)~,\\
 & & \Sigma_P = \left(\frac{m_{\eta'}}{2M}\right)^2
\left[1 + \frac{2(E +M)}{m_{\eta'}}{ 
\Pi \cdot \Pi}\right]\frac{1}{2M+m_\eta,}~.\\ 
 & & \Sigma_{\rho}^{(1)} = i \frac{m_{\eta'}}{4M^2}\nonumber\\
 & & ~~~~~~\left\{(1+\kappa \Pi \cdot \Pi)  \left[2-(1-
\frac{\kappa}{2})\frac{m_{\eta'}}{2M+m_{\eta'}} \right] +
\frac{\kappa m_{\eta'}^2}{4M(2M+m_{\eta'})} 
           \right\}~, \\
 & & \Sigma_{\rho}^{(2)} = i\kappa (1+\kappa)\frac{m_{\eta'}}{2M}
\nonumber \\
 & &  ~~~~~~\left\{-\frac{E+M}{M} {\Pi \cdot \Pi}
\left(1-\frac{m_{\eta'}}{2M+m_{\eta'}}\right) +
\frac{m_{\eta'}^2}{4M(2M+m_{\eta'})}\right\}~,\\
 & & \Sigma_{\omega}^{(1)}  
= i\frac{m_{\eta'}}{4M^2}\left(2-\frac{m_{\eta'}}{2M+m_{\eta'}}\right)~,\\
 & & \Sigma_{\omega}^{(2)} = 0~.
 \end{eqnarray}
All amplitudes are the sum of s and u nucleon pole terms, except 
$M_{\pi}$, Eqn. \ref{minpi} which includes also a $\delta$ 
meson t pole term. 

The various exchange 
contributions are shown in Fig. 5 
versus the energy available in the CM system. The solid line is 
the transition
amplitude obtained with the relative phases of all different 
exchange contributions
set to be +1. Clearly, strong cancellations amongst these
lower the transition amplitude to below $M_{\pi}$.
Most important are the $\sigma$ and $\rho$ exchanges with ratios
$M_{\sigma} : M_{\rho} : M_{\omega} : M_{\pi} : M_{\delta} : M_{\eta}
\\approx 14 : 10 : 8 : 7 : 2.5 : 1 $. 
The ratios quoted above  differ 
considerably from the ones reported in Ref. \cite{gedalin98} 
for the $\eta$ production, where the reaction proceeds mainly 
via nucleon isobar excitations and the relative importance of
various exchanges is determined by s+u nucleon isobar pole
terms. Also the kinematic $\Sigma_B$ factors in 
Eqns.  \ref{minsigma}-\ref{minomega} vary strongly with the 
mass of the meson produced, giving rise to quite different ratios 
for the s+u nucleon pole terms.
At threshold the vertices $V_{\delta \eta \pi}, (k^2 = m_{\eta}^2)$ 
and $V_{\delta \eta' \pi}, (k^2 = m_{\eta'}^2)$ are small and practically
do not affect the calculated cross sections.

Our predictions for the total cross sections are shown in Fig. 6 , along
with the data from 
Refs.\cite{hibou98,moskal98,bergdolt93,chiavassa94,calen96,
schuberth95,calen96,haggstrom97}. 
The curves shown are corrected for final state interactions
according to the procedure described in Ref.\cite{gedalin98}.
To be consistent with the OBE picture of the NN interactions 
\cite{machleidt89} we have disregarded $\eta'$ exchange contributions.
Such contributions scale like $(g_{\eta 'NN}/g_{\eta NN})^2$
and practically do not influence the cross sections presented here 
for the $pp \rightarrow pp \eta (\eta')$ reactions.

\section{Summary}
\bigskip

In summary we have applied a covariant OBE model to calculate
cross sections for the $\pi^0p  \rightarrow p \eta ' (\eta)$
and $pp \rightarrow pp \eta '(\eta)$ reactions. 
In marked difference with the $\eta$, we could reproduce near 
threshold cross sections for $\pi^0 p \rightarrow \eta'p$ and
$pp \rightarrow pp \eta '$ without a resonant production via an 
intermediate baryon resonance.
As in previous studies\cite{gedalin98,gedalin981} the model
is based on the OBE picture of the NN force\cite{machleidt89}
and accounts for relativistic effects, energy dependence and 
nonlocality of the hadronic interactions.  
The calculations reported here and in 
Refs. \cite{gedalin98,gedalin981} for $pp \rightarrow pp \pi^0 (\eta)$
provide a consistent description for  
pseudoscalar meson production in NN collisions. All calculations 
are carried out with the same formalism and meson-nucleon couplings as 
obtained by Machleidt\cite{machleidt89} from fitting NN scattering data.
The model
success to explain data for these processes depends on how well the
half off mass shell amplitude for the conversion processes,
$BN \rightarrow PN$ are calculated. As there are no data available 
which link directly to the off mass shell behavior of these amplitudes,
this feature of the model needs still further verifications. Yet it is
encouraging that such a simple model reproduce the production rate 
for as light
as a pion and as heavy as the $\eta'$ meson.
\ \\
\ \\
{\bf Acknowledgments} This work was supported in part by the Israel
ministry Of Immigration and Absorption. We would like to thank Dr. Z.
Melamed for assistance in computation.

\setlength{\oddsidemargin}{0in}
\setlength{\evensidemargin}{1.6in}
\setlength{\textwidth}{6.0in}
\setlength{\topmargin}{-.6in}
\setlength{\textheight}{9.0in}
\pagestyle{plain}
\setlength{\baselineskip} {4.0ex}

\newpage
\vspace{0.4cm}
\begin{figure}
\vspace{4.0in}
\includegraphics{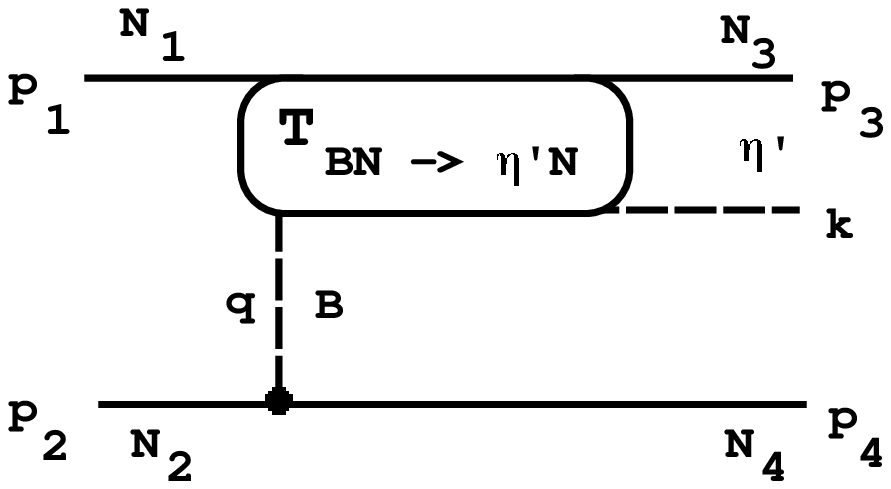}
\vskip 0.5 in
\caption{ The primary production mechanism for the 
$NN \to NN P$ reaction. A boson B created on nucleon 1 (momentum 
$p_1$), is converted into $\eta'$ (momentum k) on nucleon 2
(momentum $p_2$).  
}  
\end{figure}

\vspace{0.4cm}
\begin{figure}
\vspace{3.0in}
\includegraphics{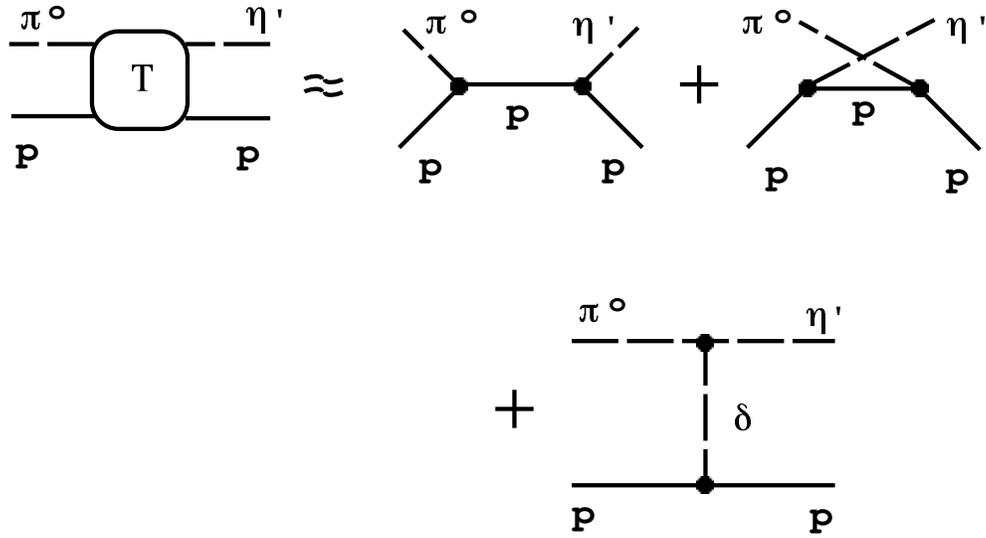}
\vskip 0.5 in
\caption{ Feynman diagrams contributing to the conversion process 
$BN \rightarrow \eta' N$. For $\eta$ production, because of the strong 
coupling to the $N^*$ (1535 MeV) nucleon isobar we include
contributions from both nucleon and nucleon isobar excitation in
the intermediate states. Graph (e) describes a $\delta$-meson pole 
in a t channel.   }  
\end{figure}

\begin{figure}[t]
\vspace{5.0in}
\includegraphics{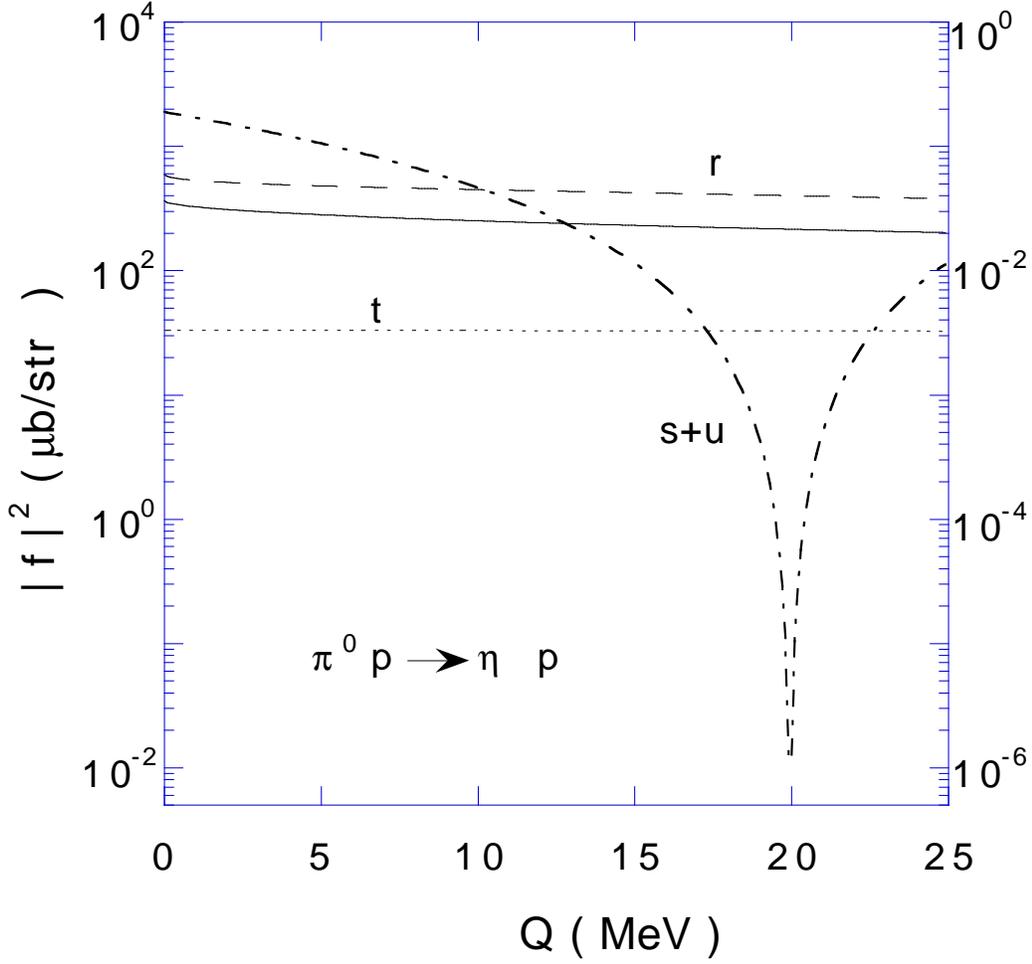}
\vskip 0.5 in
 \caption{  The  $\pi^0 p \to p \eta$ conversion amplitude :
the curves labeled by r, s+u and t represent contributions from s+u 
nucleon isobar pole terms (resonance production), s+u nucleon pole terms and
$\delta$-meson pole in a t channel, respectively (see text).
The s+u nucleon pole contributions (dash-dotted curve) are to be evaluated
using the scale on the right, while all other contributions use the scale
on the left.
}
    \label{fig3}
\end{figure}

\begin{figure}[t]
\vspace{5.0in}
\includegraphics{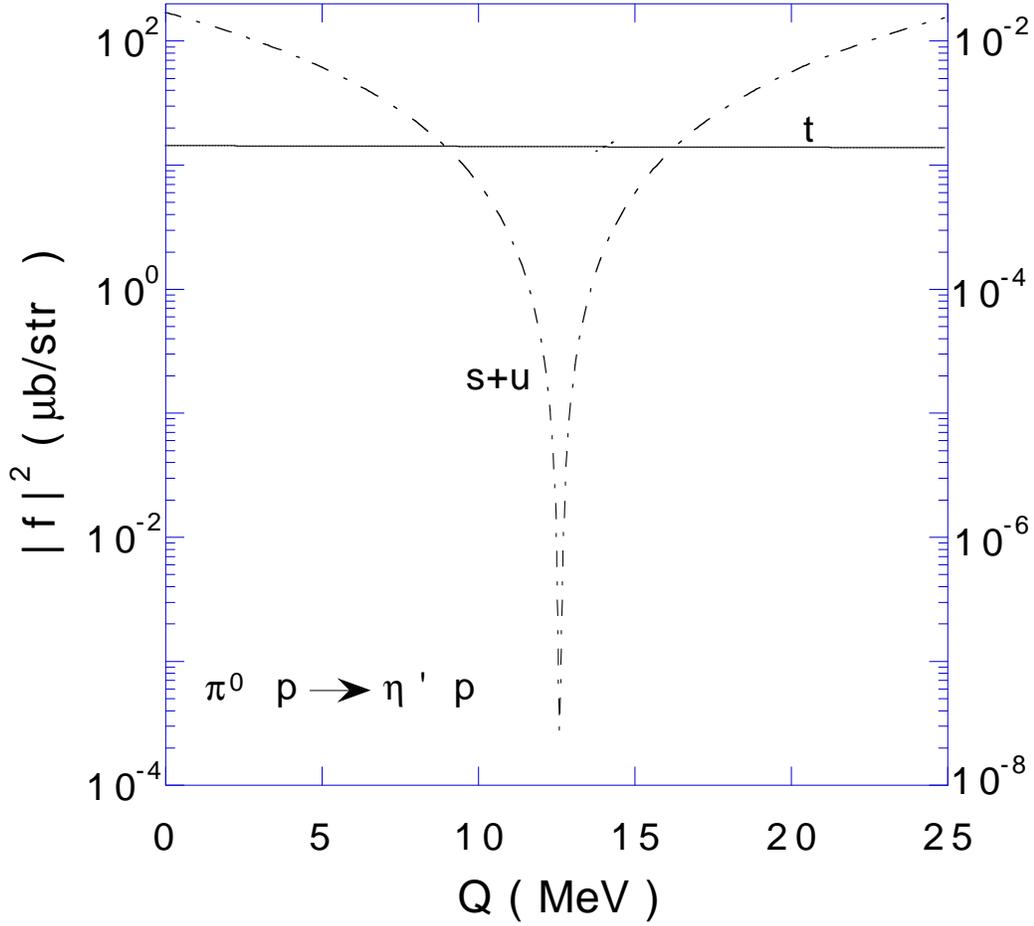}
\vskip 0.5 in
 \caption{  The  $\pi^0 p \to p \eta'$ conversion amplitude :
the curves labeled by s+u and t represent contributions 
s+u nucleon pole terms and $\delta$-meson pole in a t channel, respectively 
(see text). The t pole term contribution dominates
the process and is not resolved from the total amplitude (solid line).
The s+u nucleon pole contributions (dash-dotted curve) are to be evaluated
using the scale on the right, while the t pole contribution and the total
amplitude and total amplitude use the scale on the left.
}
    \label{fig4}
\end{figure}

\begin{figure}[t]
\vspace{4.5in}
\includegraphics{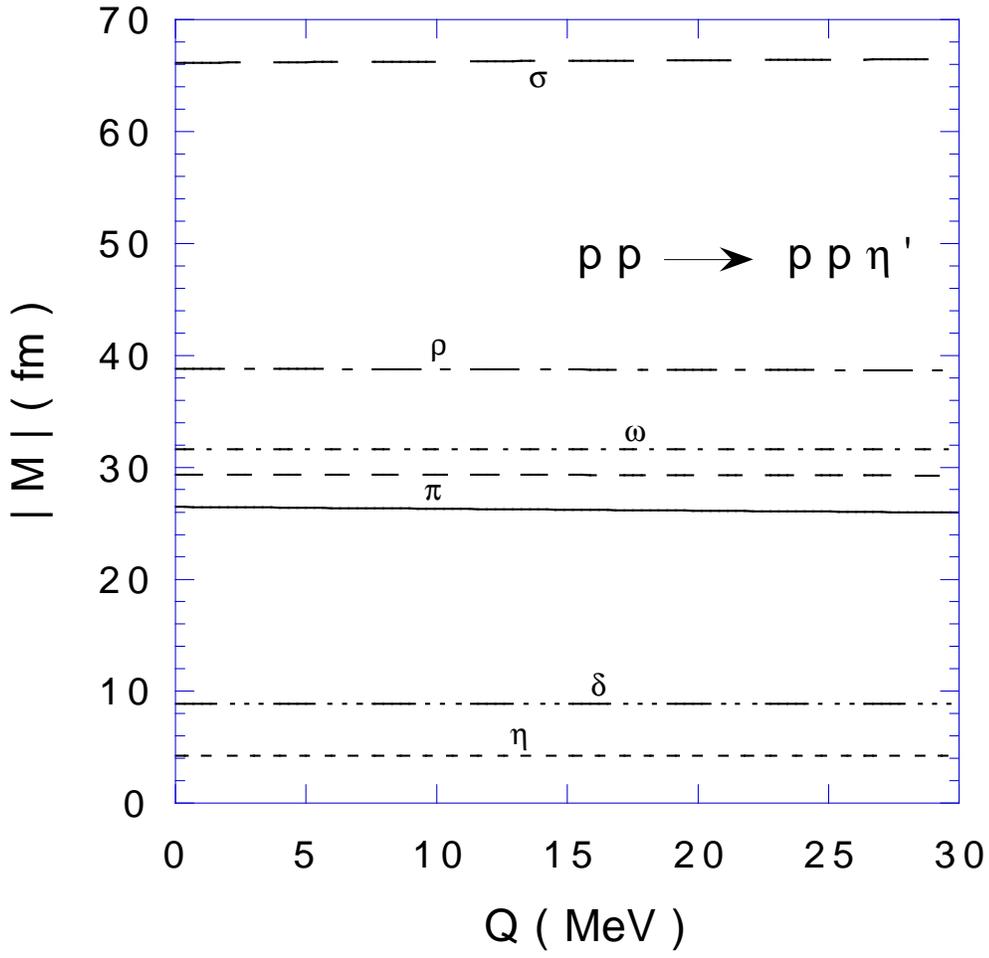}
\vskip 0.5 in
 \caption{  Partial exchange amplitude for the $pp \to pp \eta'$
reaction. The contribution from the $\delta$-meson pole is included   
in $M_{\pi}$. }
    \label{fig4}
\end{figure}

\begin{figure}[t]
\vspace{5.5in}
\includegraphics{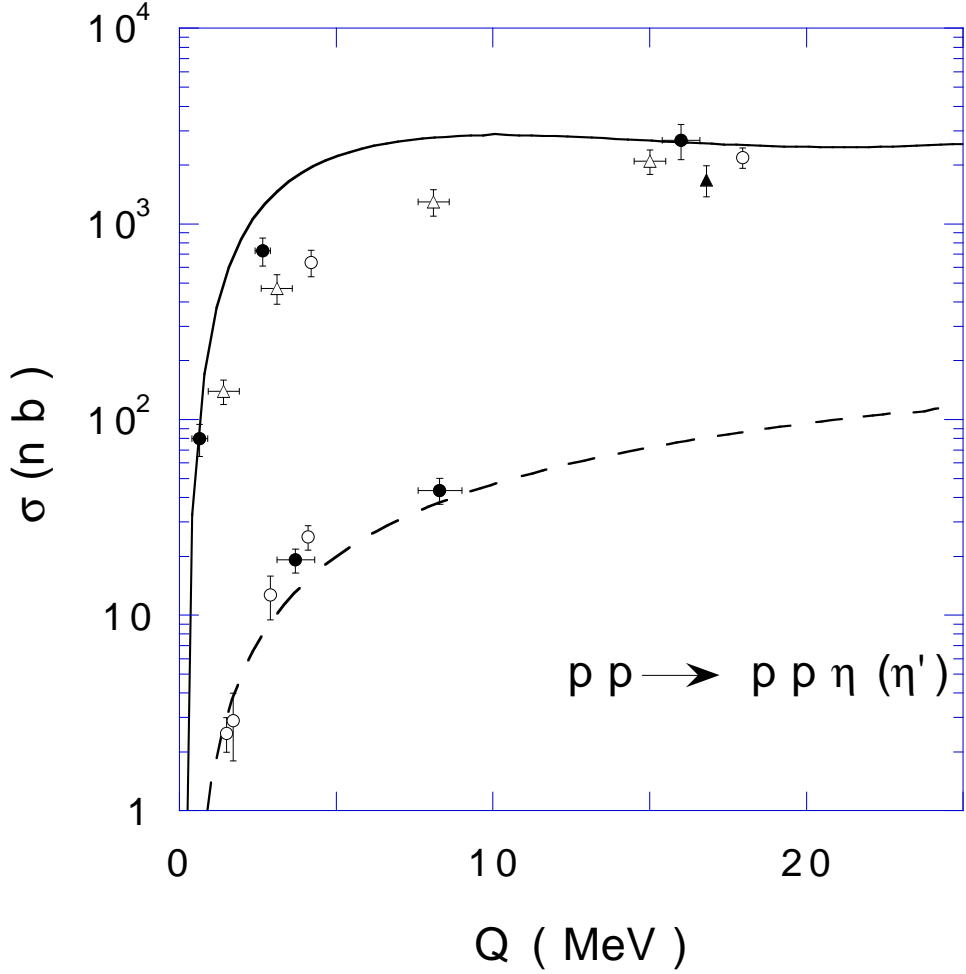}
\vskip 0.5 in
 \caption{ Energy integrated cross sections for the $pp \rightarrow pp \eta$
reaction $versus$ the energy available in the CM system. All curves include 
FSI corrections via the approximation of Ref.\cite{gedalin98}. 
The calculated cross section for the $pp \rightarrow pp \eta$ with the
$\delta$ pole term included (solid line) is practically identical with
prediction reported in Ref.\cite{gedalin98}. 
}    

\end{figure}

\end{document}